\documentclass[conference,a4paper]{IEEEtran}
\usepackage{amsmath,amsthm,amssymb}
        \usepackage{lipsum}

\newcommand\blfootnote[1]{%
  \begingroup
  \renewcommand\thefootnote{}\footnote{#1}%
  \addtocounter{footnote}{-1}%
  \endgroup
}             
\usepackage{enumerate}
\usepackage{color}
\usepackage{xcolor}
\usepackage{url}
\usepackage{balance}
\usepackage{graphicx} 
\usepackage{mathrsfs}
\usepackage{epsfig}
\usepackage{verbatim}
\usepackage{setspace}
\usepackage{cite}
\usepackage{multicol}
\usepackage{lipsum}
\usepackage{etoolbox}
\usepackage{algpseudocode}
\usepackage{algorithmicx}
\usepackage{algorithm}

\newtheorem{theorem}{Theorem}
\newtheorem{lemma}{Lemma}

\newtheorem{corollary}{Corollary}
\newtheorem{definition}{Definition}
\newtheorem{claim}{Claim}
\newtheorem{remark}{Remark}

\newcommand{\bX}{\mathbf{X}}

\newcommand{\bx}{\mathbf{x}}
\newcommand{\bY}{\mathbf{Y}}
\newcommand{\bZ}{\mathbf{Z}}
\newcommand{\by}{\mathbf{y}}
\newcommand{\bz}{\mathbf{z}}

\newcommand{\bbP}{\mathbb{P}}

\newcommand{\cX}{{\cal X}}
\newcommand{\cY}{{\cal Y}}
\newcommand{\cZ}{{\cal Z}}
\newcommand{\cS}{{\cal S}}
\newcommand{\cC}{{\cal C}}
\newcommand{\cL}{{\cal L}}
\newcommand{\cB}{{\cal B}}

\newcommand{\cP}{{\cal P}}

\newcommand{\cA}{{\cal A}}

\newcommand{\view}{V}

\newcommand{\td}[1]{\tilde{#1}}

\newcommand{\rx}{X}

\newcommand{\removed}[1]{}

\newcommand{\aky}[1]{\textcolor{red}{{\bf AKY:}#1}}

\newcommand{\gunc}[1]{Gaussian UNC[$\gamma^2$,$\delta^2$] }
\newcommand{\pgunc}[1]{Passive-Gaussian UNC[$\gamma^2$,$\delta^2$] }

\makeatletter
\patchcmd{\@IEEEeqnarray}{\relax}{\relax\intertext@}{}{}
\makeatother

\algdef{SE}[DOWHILE]{Do}{doWhile}{\algorithmicdo}[1]{\algorithmicwhile\ #1}%

\title{Wiretapped Commitment over Binary Channels}
\author{
\IEEEauthorblockN{
Anuj Kumar Yadav\IEEEauthorrefmark{1}\removed{$^{\href{www.google.com}{\includegraphics[scale=0.08]{orcid.JPG}}}$},
Manideep Mamindlapally\IEEEauthorrefmark{2}, and
Amitalok J. Budkuley \IEEEauthorrefmark{3}.
}\\

\vspace{2mm}
\IEEEauthorblockA{
\IEEEauthorrefmark{1}School of Computer and Communication Sciences, EPFL, Switzerland.\\
\IEEEauthorrefmark{2}CWI and University of Amsterdam, The Netherlands.\\
\IEEEauthorrefmark{3}Department of Electronics and Electrical Communication Engg., IIT Kharagpur, India.}
}

\removed{\author{

        	  \IEEEauthorblockN{Anuj Kumar Yadav}
    \IEEEauthorblockA{EPFL\\Switzerland \\ anuj.yadav@epfl.ch}    
		\and
    \IEEEauthorblockN{Manideep Mamindlapally}
    \IEEEauthorblockA{QuSoft, CWI and University of Amsterdam\\ The Netherlands \\ manideep.mamindlapally@cwi.nl}
		  \and 
\IEEEauthorblockN{Amitalok Budkuley}
        \IEEEauthorblockA{IIT Kharagpur\\India \\ amitalok@ece.iitkgp.ac.in}
}}
\begin{document}
\maketitle 
\blfootnote{A short version of this work has been accepted for publication at the 2024 IEEE International Symposium on Information Theory (ISIT), Athens, Greece. This extended version includes detailed proofs of theorems, discussion and open questions.}
\begin{abstract}
We propose the  problem of \emph{wiretapped commitment}, where two parties, say \textit{committer} Alice and \textit{receiver} Bob, engage in a commitment protocol using a  noisy channel as a resource, in the presence of a \emph{eavesdropper}, say Eve.
 Noisy versions of Alice's transmission over the   wiretap  channel are received at both Bob and Eve. We seek to determine the maximum commitment throughput in the presence of a eavesdropper, i.e., \emph{wiretapped commitment capacity}, where in addition to the standard security requirements for two-party commitment, one seeks to ensure that Eve doesn't learn about the commit string. 

A key interest in this work is to explore the effect of collusion  (or lack of it) between the \textit{eavesdropper} Eve and either  Alice  or Bob. Toward the same, we present results on the wiretapped commitment capacity under the so-called \emph{$1$-private regime} (when Alice or Bob cannot collude with Eve)  and the \emph{$2$-private regime} (when Alice or Bob may possibly collude with Eve). 

\end{abstract}
%
%

\section{Introduction}\label{sec:introduction}

%

A classic \emph{two-party}  primitive that finds wide application in cryptographic applications is \emph{commitment} introduced by Blum~\cite{blum}. In this work, we study a \emph{three-party variation}  of this problem, viz., \emph{wiretapped commitment}, involving \textit{committer} Alice, \textit{receiver} Bob and an \textit{eavesdropper} Eve. Imagine Alice and Bob engaging in a high-stakes business transaction, exchanging digital contract details over a communication channel that may be susceptible to wiretapping by an eavesdropper Eve. 
In this scenario, there's a concern that either Alice or Bob, individually or in possible collusion with Eve, may interact with the aim of either revealing the contract terms prematurely or altering them without detection
at the time of finalization. 
We seek to devise \emph{eavesdropper-resilient} commitment schemes, also called \emph{wiretapped  commitment schemes} henceforth, such that  Alice and Bob can establish a secure and tamper-evident commitment protocol, mitigating the risks associated with wiretapping and possible collusion. Such a scheme would not only safeguard the confidentiality of the contract terms but also maintain the integrity of the agreement in the face of potential adversarial collaboration.


While Blum's classic work introduced commitment, it also brought to the fore the limitation that information-theoretically secure schemes were impossible under entirely noiseless interactions between two-parties.\footnote{Blum's work established the possibility of \emph{conditionally-secure} commitment schemes under computationally-limited parties.} Wyner's foundational work on \emph{wiretap channels}~\cite{wyner1975wire} first demonstrated the use of \emph{noisy channels} to devise schemes with  information-theoretic security (albeit in the `weak' sense). Through a series of subsequent works~\cite{crepeau1,crepeau2}, positive rate unconditionally secure commitment was shown to be possible over binary memoryless channels. Winter \emph{et al.}~\cite{winter2003commitment} characterized the commitment capacity over any  \emph{non-redundant} discrete memoryless channel (DMC); this result was subsequently extended  to DMCs with `costs' in~\cite{mymb}. Commitment has also been explored for other channel models like the unfair noisy channels~\cite{damgaard1999possibility,crepeau2020commitment,budkuley2023possibility}, elastic channels and their cousins~\cite{khurana2016secure,budkuley2022reverse}, quantum channels~\cite{lo1998quantum}, etc. Several other variants of commitment too have been studied (see~\cite{fischlin2001trapdoor,juels1999fuzzy,bloch12}).

Closer to the theme of this work, and a key motivation for our study, is the work by Mishra \emph{et al.}~\cite{mishra2017wiretapped,manoj2} on \emph{wiretapped oblivious transfer}, where the authors studied another related cryptographic primitive called the \emph{oblivious transfer}~\cite{imai,pinto,aky} in the presence of an eavesdropper. The presence of the eavesdropper necessitated additional security requirement (w.r.t. the classical problem). Here the authors focused on certain sub-classes of binary wiretap channels with erasures and studied their capacity (with potentially \emph{honest-but-curious} users) under two versions of security guarantees: (i) the \emph{$1$-privacy} setting which precludes collusion with the eavesdropper, and (ii)  \emph{$2$-privacy} setting where collusion is possible. Inspired by their work, in our problem we study wiretap channels under similarly inspired notions of privacy (with and without colluding parties); see Section~\ref{sec:system:model} for their formal definitions.

Our work formalizes the wiretapped commitment problem and presents new results on their commitment capacity. In the following, we summarize our key contributions:\\ 

\begin{itemize}
\item We initiate a systematic study of wiretapped commitment in this work. For the $1$-privacy setting, we completely characterize the commitment capacity for the binary symmetric broadcast (BS-BC) wiretap channels  in Theorem~\ref{thm:1:priv:cap}. Here we present a  converse (upper bound) for general alphabet wiretap channels, and then specialize it for the BS-BC class of such channels, followed by a matching lower bound using a scheme tailored for the BS-BC channels. 


\item Next, we present capacity results under the more challenging $2$-privacy setting where Alice or Bob  may collude with Eve. Using a general wiretap channel converse, and constructing a specialized achievability scheme matching  that bound,  we completely characterize the $2$-private commitment capacity for the \emph{independent} BS-BC wiretap channel (see Def.~\ref{def:ibsbc}).
\end{itemize}

\indent\textbf{Organization of the paper:} In the following, we briefly present the basic notation in Section~\ref{sec:not:pre}. Following which we present our  problem setup in Section~\ref{sec:system:model}. 
The main results of this work  are presented in Section~\ref{sec:main:results}. We present some details of the converse proof and achievability in Section~\ref{sec:proofs}.
Finally, we make concluding remarks in Section~\ref{sec:conclusion}.


%
\section{Notation and Preliminaries:}\label{sec:not:pre}
We denote random variables by upper case letters (eg. $X$), their values  by lower case letters (eg., $x$), and their alphabets by calligraphic letters (eg. $\cX$). 
Random vectors and their accompanying values are denoted by boldface letters. 
For  natural number $a\in\mathbb{N}$, let $[a]:=\{1,2,\cdots, a\}$. 
%
Let $P_X$ denote the distribution of $X\in\cX$.
Distributions for multiple random variables are similarly defined. 
$P_{X|Y}$ and $[P_{X,Y}]_X$ denote the conditional distribution of $X$ (conditioned on $Y$) and the marginal distribution of $X$ (under the $P_{X,Y}$ joint distribution).
%
Given $P_\rx,Q_\rx \in \cP(\cX)$, $||P_{\rx}-Q_\rx||$ denotes their statistical distance.
%

Let random variables $X,Y\in\cX\times\cY$, where $(X,Y)\sim P_{X,Y}$. The \emph{min-entropy} of $X$ is denoted by $H_{\infty}(X):=\min_{x\in\cX} \left(-\log(P_{X}(x))\right)$; the conditional version is given by $H_{\infty}(X|Y):=\min_{y} H_{\infty}(X|Y=y).$ For $\epsilon\in[0,1)$, the \emph{$\epsilon$-smooth min entropy} and its conditional version is given by:
$H_{\infty}^\epsilon(X):= \max_{X':||P_{X'}-P_{X}||\leq\hspace{1mm}\epsilon} H_{\infty}(X')$ and 
$H_{\infty}^\epsilon(X|Y):= \max_{X',Y':||P_{X',Y'}-P_{X,Y}||\leq\hspace{1mm}\epsilon} H_{\infty}(X'|Y')$ respectively.

\removed{
Let $X \in \cX$ and $Y \in \mathbb{R}$ represent a discrete random variable and a continuous random variable respectively. For every $x \in \cX$, the conditional probability density function (PDF) $f_{Y|X}(y|x)$ is assumed to be Riemann integrable. Then, the PDF of $Y$ is given by 
\begin{align*}
    f_{Y}(y)=\sum_{x \in \cX}P_{X}(x)f_{Y|X}(y|x)
\end{align*}
is also Reimann Integrable.\\
Further, the conditional probability $P_{X|Y}(x|y)$ is given by
\begin{align*}
    P_{X|Y}(x|y)=\frac{P_{X}(x)f_{Y|X}(y|x)}{f_{Y}(y)}
\end{align*}
The conditional entropy of $X$ given the random variable $Y$ is given by:
\begin{align*}
    H(X|Y)=\int_{\infty}^{\infty}f_{Y}(y)\Bigg(\sum_{x \in \cX}P_{X|Y}(x|y)\log\frac{1}{P_{X|Y}(x|y)}\Bigg)dy
\end{align*}
}%

We also need universal hash functions and strong randomness extractors for our commitment scheme; see~\cite{nisan1996randomness,dodis2004fuzzy,bloch} for detailed definitions. 

%
\begin{definition}[$\xi$-Universal hash functions]
	Let $\mathcal{H}$ be a class of functions from $\cX$ to $\cY$. $\mathcal{H}$ is said to be $\xi-$universal hash function, where $\xi\in\mathbb{N}$, if when $h\in\mathcal{H}$ is chosen uniformly at random, then $(h(x_1),h(x_2),...h(x_{\xi}))$ is uniformly distributed over $\cY^{\xi}$, $\forall x_1,x_2,...x_{\xi} \in \cX$.
\end{definition}
\begin{definition}[Strong randomness extractors]
	 A probabilistic polynomial time function of the form \text{Ext}: $\{0,1\}^n \times \{0,1\}^d \to \{0,1\}^m$ is an  $(n,k,m,\epsilon$)-strong extractor if for every probability distribution $P_{Z}$ on $\cZ=\{0,1\}^n$, and $H_{\infty}(Z)\geq k$, for random variables $D$ (called 'seed') and $M$, distributed uniformly in $\{0,1\}^d$ and $\{0,1\}^m$ respectively, we have $||P_{Ext(Z;D),D}- P_{M,D}|| \leq \epsilon$.
\end{definition}
\section{System Model and Problem Description}\label{sec:system:model}

\begin{figure}[!ht]
  \begin{center}
		\includegraphics[trim=3cm 8cm 5cm 0cm, scale=0.3]{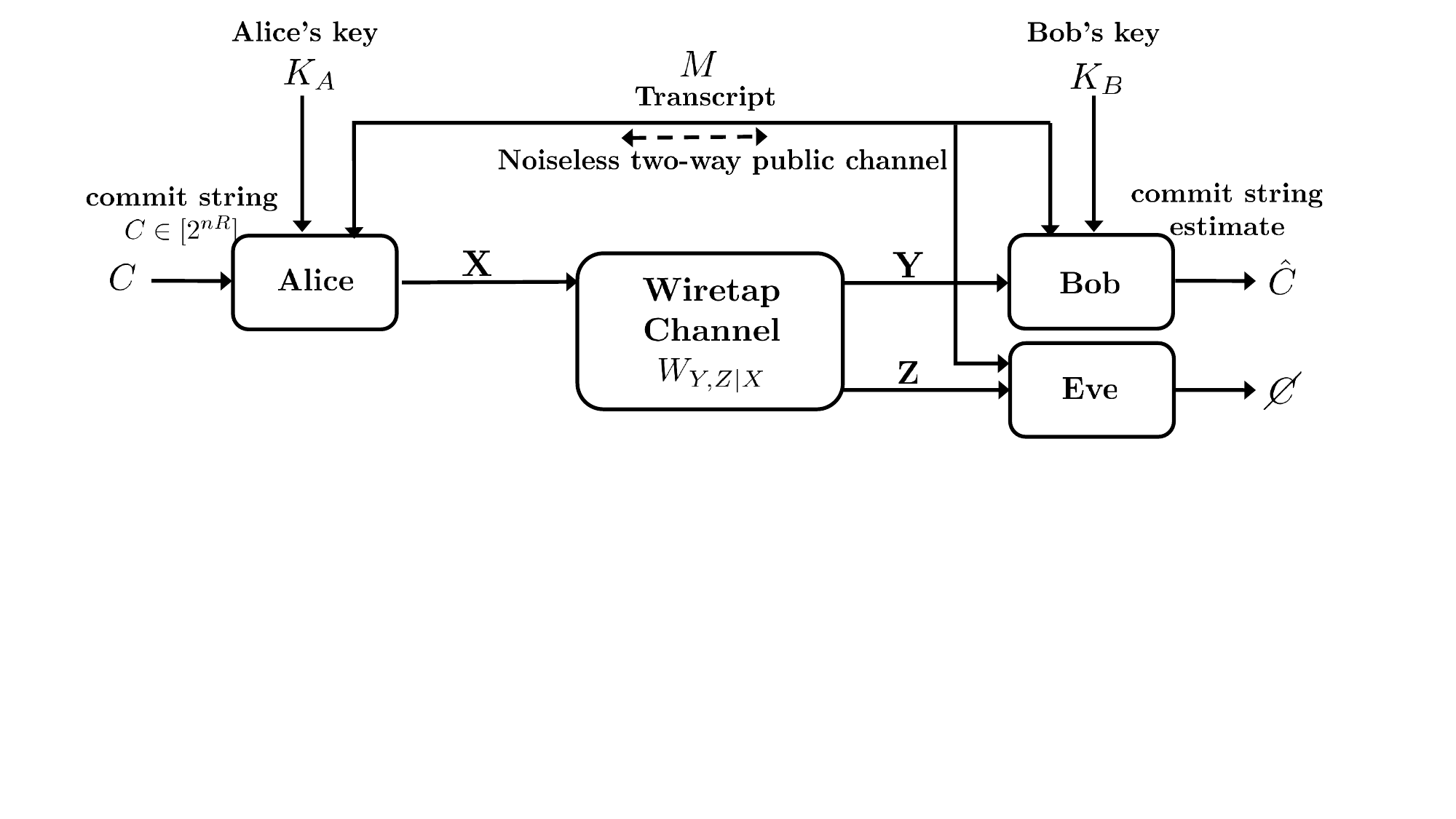}
    \caption{The problem setup: commitment over a  wiretap channel}
    \label{fig:main:setup}
  \end{center}
\end{figure}

%
Our problem (refer Fig.~\ref{fig:main:setup}) comprises three parties: \emph{committer} Alice, \emph{receiver} Bob and \emph{eavesdropper} Eve.  Alice and Bob are mutually distrustful parties and employ a noisy wiretap channel (where potentially different noisy versions of Alice's transmission are broadcast to Bob and Eve)  to realize commitment  in the presence of Eve. The commit string $C$ is chosen uniformly at random by Alice, where $C\in[2^{nR}]$, and she transmits her encrypted data or \emph{codeword} $\bX$ over a one-way  wiretap channel where channel outputs $\bY$ and $\bZ$ are received at Bob and Eve, respectively. We formally define the wiretap channel below:
\begin{definition}[Wiretap channel~\cite{wyner-inc1978},~\cite{elgamal-kim}]\label{def:wiretap:channel}
A wiretap channel is a memoryless broadcast channel with Alice's input $X\in\cX$, and outputs $Y\in\cY$ and $Z\in\cZ$ at Bob and Eve, respectively. The memoryless channel law is  given by $W_{Y,Z|X} : \mathcal{X} \to \mathcal{Y} \times \mathcal{Z}$ and is known to all parties.
\end{definition}
In this work, we specifically focus on  the following class of  binary wiretap channels.
\begin{definition}[BS-BC wiretap channels~\cite{elgamal-kim}]
A binary input binary output (BIBO) memoryless broadcast channel $W_{Y,Z|X}$, where $\cX=\cY=\cZ=\{0,1\},$ and the \emph{marginal} channel laws $W_{Y|X}:=[W_{Y,Z|X}]_{Y|X}$ and $W_{Z|X}=[W_{Y,Z|X}]_{Z|X}$ are binary symmetric channels (BSCs), say BSC($p$) and BSC($q$), where $0<p,q<1/2$, is called a binary symmetric broadcast (BS-BC) wiretap channel and denoted by BS-BC($p,q$). Note that the two BSCs may exhibit correlated behaviour. \footnote{Note that both $p$ and $q$ are strictly in the interior of the set $[0,1/2]$. We include this restriction since (information-theoretic) the commitment capacity can be easily characterized when either (or both) of $p$ and $q$ equal 0 or 1. }
\begin{remark}
    For  given $0<p,q<1/2$, BS-BC($p,q$) is not a unique channel but a class of channels. 
\end{remark}
\end{definition}
Next, we define two sub-classes of binary symmetric broadcast (BS-BC) wiretap channels which are relevant for this work.
\begin{definition}[I-BS-BC wiretap channel~\cite{elgamal-kim}]~\label{def:ibsbc}
An independent binary symmetric broadcast (I-BS-BC($p,q$)) wiretap channel is a BS-BC$(p,q)$ wiretap channel where the channel law $W_{Y,Z|X}$ can be decomposed in the following manner: $W_{Y,Z|X}=W_{Y|X}W_{Z|X}$ i.e., the Markov chain $Y-X-Z$ holds. In other words, the binary symmetric channels BSC($p$) and BSC($q$) have independent channel noise.

\end{definition}

\begin{definition}[D-BS-BC wiretap channel~\cite{elgamal-kim}]~\label{def:dbsbc}
A degraded binary symmetric broadcast (D-BS-BC($p,q$)) wiretap channel is a BS-BC$(p,q)$ wiretap channel where the channel law $W_{YZ|X}$ can be decomposed in the following manner: $W_{YZ|X}=W_{Y|X}W_{Z|Y},$ i.e., the Markov chain $X-Y-Z$ holds. In other words, the binary symmetric channel from Alice to Eve,  i.e., BSC($q$) 
is a physically degraded version of the binary symmetric channel from Alice to Bob,  i.e., BSC($p$).
\end{definition}

In addition to the noisy channel resource, as is common in such cryptographic primitives, we also assume that Alice and Bob can interact over a two-way link that is noiseless and where the interaction is public and fully authenticates the transmitting party.\footnote{Recall from earlier that under unconditionally-secure commitment, even single-bit commitment is impossible to realize under purely noiseless interactions~\cite{damgaard1999possibility}.} The \textit{eavesdropper} Eve is assumed to also have access to the interactions over the noiseless link.

To commit to her random string $C,$ Alice uses the BS-BC wiretap channel $W_{Y,Z|X}$,  $n$ times and transmits over it her encrypted string $\bX=(X_1,X_2,\cdots, X_n)\in\{0,1\}^n$. Bob and Eve, respectively, receive noisy versions $\bY\in\{0,1\}^n$ and $\bZ\in\{0,1\}^n$ of $\bX.$ Alice and Bob can both privately randomize their transmissions (over the noisy and noiseless links) via their respective keys $K_A\in\mathcal{K_A}$ and $K_B\in\mathcal{K_B}.$ At any point in time, say instant $i \in [n]$, Alice and Bob can also exchange messages over the public, noiseless link prior to transmitting $X_i$; let $M$ denote the entire collection of messages exchanged over the noiseless link. We call $M$ the \emph{transcript} of the protocol. We assume that at any point in time during the protocol, the transmissions of Alice and/or Bob can depend \emph{causally} on the information previously available to them. \\


We now formally introduce  an $(n,R)$-commitment protocol for the above setup.
\begin{definition}[Commitment protocol]\label{def:commitmentprotocol}
An $(n,R)-$commitment protocol $\mathscr{P}$ is a message-exchange procedure between Alice and Bob (in the presence of Eve) to realize commitment over the random bit string  $C\in[2^{nR}].$ Here $R$ is called the  \emph{rate} of the above  $(n,R)$-commitment protocol $\mathscr{P}.$ There are two phases to $\mathscr{P}:$ \emph{commit phase} followed by the \emph{reveal phase.} \\
\indent (a) Commit phase: Given $C\in[2^{nR}]$, Alice uses the BS-BC $(p,q)$, $n$ times to transmit $\bX$ over it. Correspondingly, Bob receives $\bY,$ and Eve receives $\bZ$ over this channel. The two parties (Alice and Bob) may also exchange messages over the noiseless link during the transmission of $\bX$. Let $M$ denote this transcript (over the noiseless link) of protocol $\mathscr{P}$ at the end of Commit phase.\footnote{We assume that the transcript may contain arbitrarily large, though finite, messages.} 
Let $V_A=(C,K_A,\bX,M), V_B=(K_B,\bY,M)$ and $V_E=(\bZ,M)$ denote, \emph{Alice's view}, \emph{Bob's view} and \emph{Eve's view } respectively which includes all the random variables and vectors known to the respective users {at the end} of the commit phase.\\
\indent (b) Reveal phase: In this phase, Alice and Bob  communicate only  over the noiseless public link and do not use the BS-BC wiretap channel \footnote{We point that in some works, unlike in this work, the noisy channel may be used to also realize the bidirectional noiseless link, and the rate calculation is, subsequently, normalized also over such channel uses. We do not study such a definition of rate in this work.}. Alice announces the commit string $\tilde{c}\in[2^{nR}]$ and $\td{\bX}\in \{0,1\}^n.$ Bob then performs a test $T(\td{c},\td{\bX},V_B)$ and either accepts (by setting $T=1$) the commit string $\td{c}$ or reject it (by setting $T=0$).
\end{definition}

In this work, we study \emph{achievability} of rate $R$ (defined later) with respect to two different notions of security metrics, viz., the $1-$privacy (where none of the legitimate parties i.e., Alice and Bob are allowed to collude with the \textit{eavesdropper} Eve) and the  $2-$privacy (where the collusion between `a \textit{malicious} Alice and Eve' or  `a \textit{malicious} Bob and Eve' is possible).
Toward defining the same, we first introduce the security metrics for any $(n,R)-$commitment protocol $\mathscr{P}.$  
\begin{definition}[$\epsilon-$sound]\label{def:soundness}
A protocol $\mathscr{P}$ is $\epsilon-$sound if for an honest Alice and  an honest Bob,
%
$\mathbb{P}\left(T(C,\bX,V_B)\neq 1\right)\leq \epsilon.$
%
\end{definition}
\begin{definition}[$\epsilon-1- $concealing]\label{def:1:concealing}
A protocol $\mathscr{P}$ is $\epsilon-1- $concealing if for an honest Alice and under any strategy of a malicious Bob,
%
$I(C;V_B) \leq \epsilon.$
%
\end{definition}
\begin{definition}[$\epsilon-1-$ binding]\label{def:1:binding}
A protocol $\mathscr{P}$ is $\epsilon-1-$binding if for an honest Bob and under any cheating function $\mathcal{A}$ of a malicious Alice,
\begin{equation*}
\mathbb{P}\Big(T(\bar{c},\bar{\bx},V_B)=1 \quad\& \quad T(\hat{c},\hat{\bx},V_B)=1\Big)\leq \epsilon\label{eq:1:binding}
\end{equation*}
for any two pairs $(\bar{c},\bar{\bx}), (\hat{c},\hat{\bx}) = \mathcal{A}(V_A)$ where $\bar{c}\neq \hat{c},$ and $\bar{\bx},\hat{\bx} \in \{0,1\}^n$.
\end{definition}
\begin{definition}[$\epsilon-$secure]\label{def:secure}
A protocol $\mathscr{P}$ is $\epsilon-$secure if for any strategy of the eavesdropper Eve,
%
$I(C;V_E) \leq \epsilon.$
%
\end{definition}
\begin{definition}[$\epsilon-2- $concealing]\label{def:2:concealing}
A protocol $\mathscr{P}$ is $\epsilon-2- $concealing if for an honest Alice and under any strategy of a malicious Bob  possibly in collusion with Eve,
%
$I(C;V_B,V_E) \leq \epsilon.$
%
\end{definition}

It might be of interest to note here that under the $2-$privacy caase, $\epsilon-2- $concealing condition directly implies that the protocol is $\epsilon-$secure as well (due to the chain rule of mutual information and the fact that mutual information is non-negative). Alternatively, we could have also defined $\epsilon-2-$secure condition which would be exactly similar to $\epsilon-2-$concealing beacause the collusion between Bob and Eve leads to both having the same view i.e., $(V_B,V_E)$.

\begin{definition}[$\epsilon-2-$binding]\label{def:2:binding}
A protocol $\mathscr{P}$ is $\epsilon-2-$binding if for an honest Bob and under any cheating function $\mathcal{A}$ of a malicious Alice possibly in collusion with Eve,
\begin{equation*}
\mathbb{P}\Big(T(\bar{c},\bar{\bx},V_B)=1 \quad\& \quad T(\hat{c},\hat{\bx},V_B)=1\Big)\leq \epsilon\label{eq:2:binding}
\end{equation*}
for any two pairs $(\bar{c},\bar{\bx}), (\hat{c},\hat{\bx}) = \mathcal{A}(V_A,V_E)$ where $\bar{c}\neq \hat{c},$ and $\bar{\bx},\hat{\bx} \in \{0,1\}^n$.
\end{definition}

A rate $R$ is said to be \emph{achievable under $1$-privacy} (resp. \textit{achievable under $2$-privacy})  if for every $\epsilon>0$ arbitrarily small and $n$ sufficiently large, there exists a commitment protocol $\mathscr{P}$ such that  $\mathscr{P}$  is $\epsilon-$sound, $\epsilon-1-$concealing  (resp. $\epsilon-2- $concealing) and $\epsilon-1- $binding (resp. $\epsilon-2-$binding) and $\epsilon-$secure.

The supremum of all achievable $1- $private rates  (resp. $2-$private rates) is called the \emph{$1- $private commitment capacity} (resp. \emph{$2- $private commitment capacity}), and the commitment capacity capacity  is denoted by $\mathbb{C}_1$ (resp. $\mathbb{C}_2$).

\section{Our Main Results}\label{sec:main:results}

%
We now state our key results in this section.
\begin{theorem}[Capacity of BS-BC under $1-$privacy]\label{thm:1:priv:cap}
    Consider a channel $W_{Y,Z|X}\in$ BS-BC$(p,q)$, where $0<p,q<1/2.$ Then, the  commitment capacity $\mathbb{C}_1$ of such a BS-BC($p,q$) under $1-$privacy is 
    \begin{align}\label{eq:1:privacy:cap}
        \mathbb{C}_1=\min\{H(p),H(q)\}.
    \end{align}    
\end{theorem}

We present the  full proof in Section~\ref{sec:proofs:c1} and Section~\ref{sec:proofs:achieve:1}. \\

We first a present a converse for a general (finite) alphabet  wiretap channel and then specialize the same for the BS-BC$(p,q)$ wiretap channel. Our achievability (inspired from~\cite{damgaard1999possibility,crepeau2020commitment,yadav2021commitment}) is tailored for the BS-BC wiretap channels and utilizes random hash exchange challenge and a strong randomness extractor based on 2-universal hash function. However, unlike any of the previous works, our scheme just requires \textit{one round} of random hash challenge essentially to guarantee bindingness.


For the  I-BS-BC($p,q$) the capacity expression remain unchanged from that in~\eqref{eq:1:privacy:cap} since that expression depends entirely on the marginal channel laws $W_{Y|X}$ and $W_{Z|X}$  (which are specified via general parameters $p$ and $q$). For the Degraded-BS-BC, however, we have the following immediate corollary.
\begin{corollary}[Capacity of D-BS-BC under $1- $privacy]
    The commitment capacity for the degraded binary symmetric broadacst wiretap channel, D-BS-BC($p,q$), under $1-$privacy  is H$(p)$.
\end{corollary}
The result follows by noting that for the D-BS-BC($p,q$), we can express  $q=p*\theta,$ for some $\theta\in[0,1/2).$ Hence, $q\geq p,$ and the minimum in~\eqref{eq:1:privacy:cap} evaluates to $H(p)$.

Next, we state our results for capacity under 2-privacy. For this setting, we completely characterize the 2-private commitment capacity for the I-BS-BCO($p,q$). We state the following theorem:
\begin{theorem}[Capacity of I-BS-BC under $2-$privacy]
    Consider an I-BS-BC($p,q$) where $0<p,q<1/2.$ Then, the commitment capacity $\mathbb{C}_2$ of such I-BS-BC($p,q$) under $2-$privacy is 
\begin{align}\label{eq:2:private:cap}
\mathbb{C}_2=H(p)+H(q)-H(p\circledast q).    
\end{align}
    where $p\circledast q$ denotes the binary convolution of $p$ and $q$ i.e., $p \circledast q:= p(1-q)+q(1-p)$.
\end{theorem}
We present the proof in Section~\ref{sec:proofs}. 
 For the converse, we present a rate upper bound, viz., $R\leq H(X|Y,Z),$ for any general wiretap channel. This bound is then specialized to the binary setting (for I-BS-BC($p,q$)) and evaluates to the expression given in Eq~\eqref{eq:2:private:cap}. It is pertinent to note that the above bound holds for \emph{any} wiretap channel. For the I-BS-BC specifically, we present a matching lower bound through an achievable scheme. It has similarities to the scheme in the 1-private setting but crucially differs in the choice of the hash families and the randomness extractor.  


%

\section{Proofs}\label{sec:proofs}

\removed{\aky{This section includes the following:
\begin{itemize}
    \item 1-privacy converse for general alphabet wiretapped bradcast channnels.
    \item Specialised the above converse for the binary case wiretapped channel where the channel matrices $P_{Y|X}$ and $P_{Z|X}$. This channel model includes the independent as well as the degraded case.
    \item 1-privacy achievability for binary case wiretapped broadcast channel. The converse and achievability together show that the capacity of binary wiretapped broadcast channnels is $\min\{H(p),H(q)\}$.
    \item Then, we have the $2-$privacy converse which gives an upper bound on the capacity of a general alphabet wiretapped broadcast channel under 2 privacy. We specialise this converse to a binary case `but inpendent' wiretapped broadcast channel. Note that this upper bound also works for the degraded case its just that the upper bound might be loose. (though, we need to compute it for it.)
    \item Then, we have a $2-$privacy achievability that only works for the binary case independent wiretapped broadcast channels. They together show that the capacity under 2-privacy is known only for binary case independent wiretapped broadcast channels.
\end{itemize}
}}
\subsection{$1$-privacy converse analysis for the wiretap channel $W_{Y,Z|X}$}\label{sec:proofs:c1}
In this subsection, we will first derive an upper bound on the rate of any commitment protocol for a general wiretap channel $W_{Y,Z|X}$ under $1$-privacy. Then, we specialize the result to get a tight upper bound for the binary symmetric broadcast (BS-BC$(p,q)$) wiretap channel. Additionally, we \emph{strengthen} our converse by proving that our upper bound on the commitment rate holds even under a \emph{weaker} notion of $\epsilon$-$1$-concealment and $\epsilon$-secrecy against Eve, which are defined below:\footnote{This is a security notion directly inspired from the weak secrecy metric  originally studied by Wyner~\cite{wyner1975wire} for wiretap channels. 
}
\begin{definition}[$\epsilon$-weakly-$1$-concealing]
An $(n,R)$-commitment protocol is said to be $\epsilon$-weakly-$1$-concealing  if for an honest Alice and under \emph{any} strategy of Bob,
\begin{align}
\frac{1}{n}I(C;\view_B)\leq \epsilon.\label{eq:ws:1}
\end{align}
\end{definition}
\begin{definition}[$\epsilon$-weakly-secure]
An $(n,R)$-commitment protocol is said to be $\epsilon$-weakly-$1$-secure against the eavesdropper Eve if under \emph{any} strategy of Eve,
\begin{align}
\frac{1}{n}I(C;\view_E)\leq \epsilon.\label{eq:ws:1}
\end{align}
\end{definition}
Now, consider any sequence of commitment protocols $(\mathscr{P}_n)_{n\geq 1}$, such that $\forall n$, $\mathscr{P}_n$ is $\epsilon_n$-sound, $\epsilon_n$-weakly-$1$-concealing, $\epsilon_n$-$1$-binding, and $\epsilon$-weakly-secure, such that $\epsilon_n\geq 0$ and $\epsilon_n\rightarrow  0$ as $n\rightarrow \infty$.

For these sequence of protocols, we  state the following lemma which upper bounds the conditional entropy $\frac{1}{n}H(C|\bX,V_{B})$ using Fano's inequality; we will use this lemma  later  to upper bound the commitment rate.
\begin{lemma}\label{lem:H:C:XV}
For every commitment protocol $\mathscr{P}_n$ satisfying all the security guarantees under $1-$privacy, we have $\frac{1}{n}H(C|\bX,V_{B})\leq \epsilon''_n$, where $\epsilon''_n\rightarrow 0$ as $n\rightarrow \infty$.
\end{lemma}
The proof appears in Appendix~\ref{app:A}, and follows from the fact that each protocol $\mathscr{P}_n$ satisfies $\epsilon_n-$soundness and $\epsilon_n-$bindingness. 

Let us now bound the commitment rate $R$ as follows:
\begin{align}
R&\stackrel{}{=} \frac{1}{n}H(C)\notag\\
&\stackrel{(a)}{=} \frac{1}{n}H(C|V_B)+ \frac{1}{n}I(C;V_B)\notag\\
	&\stackrel{(b)}{\leq} \frac{1}{n}H(C|V_B)+ \epsilon_n\notag\\
	&\stackrel{(c)}{=} \frac{1}{n}H(C|{\bY}_{}, K_B, M) + \epsilon_n\notag\\
	&\stackrel{(d)}{\leq} \frac{1}{n}H(C,\bX|{\bY}_{}, K_B, M) + \epsilon_n\notag\\
	&\stackrel{(e)}{=} \frac{1}{n}H(\bX|{\bY}_{}, K_B, M) + \frac{1}{n}H(C|\bX,{\bY}_{}, K_B, M)+\epsilon_n\notag\\
	&\stackrel{}{=} \frac{1}{n}H(\bX|{\bY}_{}, K_B, M) + \frac{1}{n}H(C|\bX,V_B)+\epsilon_n\notag\\
	&\stackrel{(f)}{\leq} \frac{1}{n}H(\bX|{\bY}) + \epsilon''_n+\epsilon_n\notag\\
	&\stackrel{}{\leq} \frac{1}{n}\sum_{i=1}^n H(X_i|Y_{i}) + \epsilon''_n+\epsilon_n\notag \\
    &\stackrel{}{\leq}\max_{P_{X}}H(X|Y)+\epsilon''_n+\epsilon_n\label{eq:rate:11}
\end{align}
Here, 
\begin{enumerate}[(a)]
\item follows from the definition of mutual information.
\item follows from the fact that every commitment protocol in the sequence $\mathscr{P}_n$ is $\epsilon_n$-weakly-$1$-concealing.
\item follows from noting that Bob's view $V_B=(\bY,M,K_B),$ where $\bY$ denotes the output of the wiretap channel.
\item follows from property of joint entropy.
\item follows from the chain rule of joint entropy.
\item follows from lemma~\ref{lem:H:C:XV} and noting that conditioning reduces entropy. 
\end{enumerate}
Therefore, finally we have:
\begin{align}\label{eq:R:3}
	R&\stackrel{}{\leq} \max_{P_{X}} H(X|Y)
\end{align}

Let $W_{Y|X}:=[W_{Y,Z|X}]_{Y|X}$ and $W_{Z|X}:=[W_{Y,Z|X}]_{Z|X}$ denote the effective channels from Alice to Bob and Alice to Eve, respectively. Suppose there exists a channel $U_{\tilde{Y}|Z}: \mathcal{Z}\rightarrow \mathcal{Y}$ s.t.
\begin{align}\label{eq:c222}
    W_{Y|X}=W_{Z|X}U_{\tilde{Y}|Z}, 
\end{align}
\footnote{Remember that $W_{Z|X}U_{\tilde{Y}|Z}$ represents composition of $W_{Z|X}$ and $U_{\tilde{Y}|Z}$. (c.f. definition \ref{def:wiretap:channel}).} This happens when the channel from Alice to Bob is a \textit{stochastically degraded} version of the channel from Alice to Eve. Consider a \textit{cheating strategy} adopted by the \textit{eavesdropper} Eve to learn about the commit string $c$. The eavesdropper Eve passes the received vector $\bZ$ \textit{locally} through the simulated \textit{private channel} $U_{\tilde{Y}|Z}$ to generate a $\tilde{\bY}$ which is a candidate $\bY$ i.e., $P_{X,\tilde{Y}}=P_{X}W_{Y|X}=P_{XY}$. Note that the view of Eve here is $V_E=(\bZ,\tilde{\bY},M)$.

Suppose such a $U_{\tilde{Y}|Z}$ exists. Then, from Lemma~\ref{lem:H:C:XV}, we have that
\begin{align}
    H(C|\bX,\bY,K_B,M) \leq \epsilon''_n \notag\\
    H(C|\bX,\bY,M) \stackrel{(a)}{\leq} \epsilon_n'' \notag\\
    H(C|\bX,\tilde{\bY},M)\stackrel{(b)}{\leq}\epsilon''_n \notag\\
     H(C|\bX,Z,\tilde{\bY},M)\stackrel{(c)}{\leq}\epsilon''_n \notag\\
      H(C|\bX,V_E)\stackrel{}{\leq}\epsilon''_n \label{eq:nc}
\end{align}
where,
\begin{enumerate}[(a)]
    \item follows from noting that $K_B$ just models local randomness at Bob which is independent.
    \item follows from the assumption of the markov chain $M-\bX-\bY$ (see Remark~\ref{rem:aa:1}), and thus noting that the joint distributions of $(C,\bX,\bY,M)$ and $(C,\bX,\tilde{\bY},M)$ are equal.
    \item follows from the fact that conditioning reduces entropy.
\end{enumerate}

Now, we can also upper bound the commitment rate as follows:

\begin{align}
R&\stackrel{}{=} \frac{1}{n}H(C)\notag\\
&\stackrel{}{=} \frac{1}{n}H(C|V_E)+ \frac{1}{n}I(C;V_E)\notag\\
	&\stackrel{(a)}{\leq} \frac{1}{n}H(C|V_E)+ \epsilon_n\notag\\
 &\stackrel{(b)}{\leq} \frac{1}{n}H(C,\bX|V_E)+ \epsilon_n\notag\\
	&\stackrel{(c)}{=} \frac{1}{n}H(\bX|{\bZ}_{},\tilde{\bY}, M) + \frac{1}{n}H(C|\bX,{\bZ}_{}, \tilde{\bY}, M)+\epsilon_n\notag\\
	&\stackrel{}{=} \frac{1}{n}H(\bX|{\bZ}_{},\tilde{\bY}, M) + \frac{1}{n}H(C|\bX,V_E)+\epsilon_n\notag\\
	&\stackrel{(d)}{\leq} \frac{1}{n}H(\bX|{\bZ}) + \epsilon''_n+\epsilon_n\notag\\
	&\stackrel{}{\leq} \frac{1}{n}\sum_{i=1}^n H(X_i|Z_{i}) + \epsilon''_n+\epsilon_n\notag \\
    &\stackrel{}{\leq}\max_{P_{X}}H(X|Z)+\epsilon''_n+\epsilon_n\label{eq:rate:11}
\end{align}
Here, 
\begin{enumerate}[(a)]
\item follows from the fact that every commitment protocol in the sequence $\mathscr{P}_n$ is $\epsilon_n$-weakly-secure against any statregy of Eve.
\item follows from the chain rule of joint entropy.
\item follows from noting that Eve's view $V_E=(\bZ,\tilde{\bY},M)$.
\item follows from the fact that conditioning reduces entropy and from Eq.~\eqref{eq:nc}.
\end{enumerate}
Finally, we also have:
\begin{align}\label{eq:R:4}
	R&\stackrel{}{\leq} \max_{P_{X}} H(X|Z)
\end{align}

The above upper bound holds only if a $U_{\tilde{Y}|Z}$ satisfying Eq.~\eqref{eq:c222} exists.

\begin{remark}\label{rem:aa:1} Note that the above approach includes a natural valid assuption of $M-\bX-\bY$ i.e., the transcript of the public communication $M$ and $\bY$ are independent given $\bX$. Most of the commitment protocols including the one presented below in our Achievability subsection follows this markov chain. This assumption was aslo made by Crepeau et.al~\cite{crepeau2020commitment} for proving the upper bound on the commitment rate for unfair noisy channels. As of now, it seems challenging to come up with a general converse approach without the above markov chain assumption, and is an interesting open problem. (see Lemma $5.1$ and Final remarks in \cite{crepeau2020commitment}, for details).
\end{remark}

From Eq. (\ref{eq:R:3}) and Eq. (\ref{eq:R:4}), we have the following upper bound on the commitment rate. 
\begin{align}\label{eq:R:5}
	R&\stackrel{}{\leq} \min\left\{\max_{P_{X}} H(X|Y),\max_{P_{X}} H(X|Z)\right\}
\end{align}


\removed{
{\color{blue} We would like to highlight here that the above proof approach for the upper bound $R \leq \max_{P_X}H(X|Z)$ under $1-$privacy differs slightly from the proof provided by us in the shorter $5-$page version. The proof in the $5-$page version has a very strong restriction that it works only under the  condition that the mapping of the commit string $c$ to $\bX$ at the Alice's encoder   is invertible. However, here we provided a more general proof under no such restriction at Alice's encoder and obtained the same upper bound. Thus, our result in Theorem~\ref{thm:1:priv:cap} for BS-BC wiretap channels remains the same. } \\
}

Thus, we have the following upper bounds on the commitment rate for a wiretap channel $W_{Y,Z|X}$:
\begin{enumerate} [(i)]
    \item If Alice to Bob channel i.e., $W_{Y|X}$ is a \textit{stochastically degraded} version of the channel from Alice to Eve i.e., $W_{Z|X}$, then
    \begin{align}\label{eq:R:55}
	R&\stackrel{}{\leq} \min\left\{\max_{P_{X}} H(X|Y),\max_{P_{X}} H(X|Z)\right\}
\end{align}
\item Else,
\begin{align}\label{eq:R:6}
    R\leq \max_{P_X}H(X|Y)
\end{align}
\end{enumerate}

\removed{Now, from Eq. (\ref{eq:R:3}) and Eq. (\ref{eq:R:4}), we have the following upper bound on the commitment rate when . 
\begin{align}\label{eq:aR:55}
	R&\stackrel{}{\leq} \min\left\{\max_{P_{X}} H(X|Y),\max_{P_{X}}. H(X|Z)\right\}
\end{align}
}

On solving the Eq.~\eqref{eq:R:5} or collectively solving the Eqs.~\eqref{eq:R:55} and ~\eqref{eq:R:6} for the binary symmetric broadcast (BS-BC$(p,q)$) wiretap channels, we have the similar following upper bound on the commitment rate for BS-BC$(p,q)$ under $1-$privacy: 
\begin{align}\label{eq:R:604}
	R &\stackrel{}{\leq} \min\big\{H(p),H(q)\big\}
\end{align}
where we observe that the optimizing the input distribution is $X\sim \text{Bernoulli} (1/2)$.

Here, we have the following: when Alice to Bob channel is \textit{stochastically degraded} version of Alice to Eve channel (i.e., $q<p$), the Eq.~\eqref{eq:R:55} evalautes to $R \leq {H(q)}$. Otherwise (i.e., $q \geq p$), the Eq.~\eqref{eq:R:6} evalautes to $R \leq {H(p)}$. Together, these reduce to the expression in the Eq.~\eqref{eq:R:604}.

This completes our converse analysis for BS-BC$(p,q)$ wiretap channels under $1-$privacy.

\subsection{$2$-privacy converse analysis for the wiretap channel $W_{Y,Z|X}$.}

In this subsection, we will first derive an upper bound on the rate of any commitment protocol for a general wiretap channel $W_{Y,Z|X}$ under $2$-privacy. Then, we specialize the result to get an upper bound for the Independent binary symmetric broadcast (I-BS-BC$(p,q)$) wiretap channel.

Additionally, similar to the previous case we \emph{strengthen} our converse by proving that our upper bound on the commitment rate holds even under a \emph{weaker} notion of $\epsilon$-$2$-concealment (defined below) and $\epsilon$-secrecy against Eve.

\begin{definition}[$\epsilon$-weakly-$2$-concealing]
An $(n,R)$-commitment protocol is said to be $\epsilon$-weakly-$2$-concealing  if for an honest Alice and under \emph{any} strategy of colluding Bob and Eve,
\begin{align}
\frac{1}{n}I(C;\view_B,\view_E)\leq \epsilon.\label{eq:ws:1}
\end{align}
\end{definition}
Now, consider any sequence of commitment protocols $(\mathscr{P}_n)_{n\geq 1}$, such that $\forall n$, $\mathscr{P}_n$ is $\epsilon_n$-sound, $\epsilon_n$-weakly-$2$-concealing, $\epsilon_n$-$2$-binding, and $\epsilon$-weakly-secure, such that $\epsilon_n\geq 0$ and $\epsilon_n\rightarrow  0$ as $n\rightarrow \infty$.

For these sequence of protocols, we  state the following lemma which upper bounds the conditional entropy $\frac{1}{n}H(C|\bX,V_{B},V_E)$ using Fano's inequality; similar to the $1$-privacy case, we will use this lemma  later  to upper bound the commitment rate.
\begin{lemma}\label{lem:H:C:XV11}
For every commitment protocol $\mathscr{P}_n$ satisfying all the security guarantees under $2-$privacy, we have $\frac{1}{n}H(C|\bX,V_{B},V_E)\leq \epsilon''_n$, where $\epsilon''_n\rightarrow 0$ as $n\rightarrow \infty$.
\end{lemma}
The proof appears in Appendix~\ref{app:A}, and follows from the fact that each protocol $\mathscr{P}_n$ satisfies $\epsilon_n-$soundness and $\epsilon_n-$2$-$bindingness. \\

Let us now bound the commitment rate $R$ as follows:
\begin{align}
R&\stackrel{}{=} \frac{1}{n}H(C)\notag\\
&\stackrel{(a)}{=} \frac{1}{n}H(C|V_B,V_E)+ \frac{1}{n}I(C;V_B,V_E)\notag\\
	&\stackrel{(b)}{\leq} \frac{1}{n}H(C|V_B,V_E)+ \epsilon_n\notag\\
	&\stackrel{(c)}{=} \frac{1}{n}H(C|{\bY}_{},\bZ, K_B, K_E, M) + \epsilon_n\notag\\
	&\stackrel{(d)}{\leq} \frac{1}{n}H(C,\bX|{\bY}_{}, \bZ, K_B, K_E, M) + \epsilon_n\notag\\
	&\stackrel{(e)}{=} \frac{1}{n}H(\bX|{\bY}_{}, \bZ, K_B, K_E, M) \notag \\ &\hspace{10mm}+ \frac{1}{n}H(C|\bX,{\bY}_{}, \bZ, K_B, K_E, M)+\epsilon_n\notag\\
	&\stackrel{}{=} \frac{1}{n}H(\bX|{\bY}_{}, \bZ, K_B, K_E, M) + \frac{1}{n}H(C|\bX,V_B, V_E)+\epsilon_n\notag\\
	&\stackrel{(f)}{\leq} \frac{1}{n}H(\bX|{\bY,\bZ}) + \epsilon''_n+\epsilon_n\notag\\
	&\stackrel{}{\leq} \frac{1}{n}\sum_{i=1}^n H(X_i|Y_{i},Z_{i}) + \epsilon''_n+\epsilon_n\notag \\
    &\stackrel{}{\leq}\max_{P_{X}}H(X|Y,Z)+\epsilon''_n+\epsilon_n\label{eq:rate:11}
\end{align}
Here, 
\begin{enumerate}[(a)]
\item follows from the definition of mutual information.
\item follows from the fact that every commitment protocol in the sequence $\mathscr{P}_n$ is $\epsilon_n$-weakly-$2$-concealing.
\item follows from noting that the collective view of colluding Bob and Eve is $(V_B,V_E)=(\bY,\bZ, K_B, K_E, M)$. 
\item follows from property of joint entropy.
\item follows from the chain rule of joint entropy.
\item follows from lemma~\ref{lem:H:C:XV11} and noting that conditioning reduces entropy. 
\end{enumerate}
Therefore, finally we have the following upper bpund on the commitment rate for a wiretapped broadcast channel $W_{Y,Z|X}$ under $2-$privacy:
\begin{align}\label{eq:R:33}
	R&\stackrel{}{\leq} \max_{P_{X}} H(X|YZ)
\end{align}
 
On solving Eq.~\eqref{eq:R:33} for the independent binary symmetric broadcast (I-BS-BC$(p,q)$) wiretap channels, we have the following upper bound on the commitment rate for I-BS-BC$(p,q)$ under $2-$privacy: 
\begin{align}\label{eq:R:64}
	R &\stackrel{}{\leq} H(p)+H(q)-H(p\circledast q)
\end{align}
where $p\circledast q:=p(1-q)+(1-p)q$, $\forall p,q\in[0,1]$ is the binary convolution between $p$ and $q$, and we observe that the optimizing the input distribution is $X\sim \text{Bernoulli} (1/2)$. This completes our converse analysis.

\subsection{Achievability for the BS-BC$({p,q})$ wiretap channel under $1-$privacy. }\label{sec:proofs:achieve:1}

\noindent Our achievability protocol is inspired by works in~\cite{damgaard1999possibility,crepeau2020commitment,yadav2021commitment} which utilize random hash exchange challenges and a strong randomness extractor based on 2-universal hash functions. However, unlike previous works, our scheme just requires \textit{one round of random hash challenge} essentially to \emph{bind} Alice to her choice in the commit phase thereby ensuring Bob's test $T$ can detect any cheating attempt by a malicious Alice during the reveal phase. 
The strong randomness extractor `Ext' extracts a secret key $\text{Ext}(\bX)$ with $nR$ nearly random bits from $\bX$. (note that the leftover hash lemma~\cite{glh} allows us to quantify the size of this key). This secret key is then XOR-ed with the commit string $c$ to realize a \emph{one-time pad} scheme, which \textit{conceals} the committed string against a malicious Bob and the wiretapper Eve in the commit phase.\\

Here are the details of our protocol.  The rate $R:=\min\{H(p),H(q)\}- \beta_2$, where the choice of $\beta_2>0$ is specified later. 
Let $\mathcal{G}:=\{g:\{0,1\}^n \rightarrow \{0,1\}^{n\beta_1}\}$ be a $2-$universal hash family, where $\beta_2>\beta_1>0$ is a small enough constant.
Further, let $\mathcal{E}:=\{\text{ext}:\{0,1\}^n \rightarrow \{0,1\}^{nR}\}$ be a $2-$universal hash family. \footnote{Note that  $R$ can be made arbitrarily close to $\mathbb{C}.$}\\

We now describe the commit and reveal phases:\\

\indent $\bullet$ \emph{Commit Phase:}  
To commit string $c\in[2^{nR}]$, the protocol proceeds as follows:\\

\noindent (C1). Given $c$, Alice sends $\bX\sim \text{Bernoulli}(1/2)$ independent and identically distributed (i.i.d.)  over the BS-BC$({p,q})$ wiretap channel; Bob receives $\bY$ while Eve receives $\bZ$. \\

\noindent (C2). Bob chooses a hash function $G\sim \text{Unif}\left(\mathcal{G}\right)$, and sends the description of $G$ to Alice over the noiseless channel.\\ 

\noindent (C3). Alice computes $G(\bX)$ and sends it to Bob over the noiseless channel.\\

\noindent (C4). Alice chooses an extractor function $\text{Ext}\sim\text{Unif}\left(\mathcal{E}\right)$ and sends $Q = c \oplus \text{Ext}(\bX)$ and the description of $\text{Ext}$ to Bob i.e., $(Q,\text{Ext})$ over the noiseless link.\footnote{In the following expression, operator $\oplus$ denotes component-wise XOR.}\\

\indent $\bullet$ \emph{Reveal Phase:} Alice proceeds as follows: \\

\noindent {(R1). Having received $\bY=\by$, Bob creates list $\cL(\by)$ of vectors given by:\footnote{Here the parameter $\alpha_1>0$ is chosen appropriately small.}
\begin{IEEEeqnarray*}{rCl}
\cL(\by):=\{\bx\in \{0,1\}^n: n(p -\alpha_1) \leq d_H(\bx,\by) \leq n(p +\alpha_1) \}.
\end{IEEEeqnarray*}
}

\noindent (R2). Alice announces  $(\td{c},\td{\bx})$ to Bob over the noiseless link.\\

\noindent (R3). Bob accepts $\td{c}$ if all the following three conditions are satisfied: $(i)$ $\td{\bx}\in\cL(\by)$, $(ii)$ $g(\tilde{\bx})=g({\bx})$, and $(iii)$ $\tilde{c}=q~\oplus \text{ext}(\tilde{\bx})$. Else, he rejects $\td{c}$ and outputs `0'.\\

Note that the wiretapper Eve also has the access to the noiseless channel so it observes everything shared between Alice and Bob over the noiseless channel during the commit as well as the reveal phase.

We now analyse and prove the security guarantees under $1-$privacy in detail for the $(n,R)$-commitment scheme, defined above:
%

[1] \underline{\emph{$\epsilon-$sound:}} For our protocol to be $\epsilon$-sound, it is sufficient to show that  $\bbP\left(\bX\not\in \cL(\bY)\right)\leq \epsilon$ when both the parties, Alice and Bob, are honest; the proof for the same follows from classic Chernoff bounds. \\

[2] \underline{\emph{$\epsilon-1-$concealing \& $\epsilon-$secure:}}
It is known that a positive rate commitment protocol is $\epsilon-1-$concealing and $\epsilon-$secure \footnote{where $\epsilon>0$ is \emph{exponentially decreasing} in blocklength $n$.} if it satisfies the notion of \emph{capacity-based secrecy} (cf.~\cite[Def.~3.2]{damgard1998statistical}) i.e., $I(C;V_B) \leq \epsilon$ and $I(C;V_E) \leq \epsilon$, respectively and vice versa. We use a well established equivalence relation between \emph{capacity-based secrecy} and the \emph{bias-based secrecy} (cf.~\cite[Th.~4.1]{damgard1998statistical}) to prove that our protocol is $\epsilon$-concealing. 

  To begin,  we  prove that our protocol satisfies bias-based secrecy by essentially proving the perfect secrecy of the key $\text{Ext}(\bX)$; here we crucially use the  \emph{leftover hash} lemma. Several versions of this lemma exists (cf.~\cite{impagliazzo1989pseudo,glh} for instance); we use the following (without proof):
\begin{lemma}[Leftover hash lemma]\label{lem:glh}
Let $\mathcal{G}=\{G:\{0,1\}^n\rightarrow \{0,1\}^l\}$ be a family of universal hash functions. Then, for any hash function $G$ chosen  uniformly at random from $\mathcal{G}$, and $W$
\begin{align*}
    \|(P_{G(W),G}-P_{U_l,G})\| \notag&\leq \frac{1}{2}\sqrt{2^{ -H_{\infty}(W)} 2{^l}}
\end{align*}
where $U_l\sim\text{Unif}\left(\{0,1\}^l\right).$
\end{lemma}
%
%
We begin by establishing the following lower bounds in Lemma~\ref{lem:smooth:min:entropy} and Lemma~\ref{lem:smooth:min:entropy2} which quantify the left-over uncertainity in $\bX$, after the information about $\bX$ is lost to Bob and Eve due to the access of $(\bY,G,G(\bX))$ and $(\bZ,G,G(\bX))$, respectively:
\begin{lemma}\label{lem:smooth:min:entropy}
For any $\epsilon_1>0, \zeta_1>0$ and $n$ sufficiently large, 
\begin{align}
	H_{\infty}^{\epsilon_1}&(\bX|\bY, G, G(\bX))\notag\\
	&\stackrel{}{\geq} { n(H(p)-\zeta_1-\beta_1)}-\log(\epsilon_1^{-1})\label{eq:h:inf:1}
    \end{align}
    The proof appears in Appendix~\ref{app:smooth:min:entropy}. 

\end{lemma}
\begin{lemma}\label{lem:smooth:min:entropy2}
For any $\epsilon_1>0, \zeta_2>0$ and $n$ sufficiently large, 
\begin{align}
	H_{\infty}^{\epsilon_1}&(\bX|\bZ, G, G(\bX))\notag\\
	&\stackrel{}{\geq} { n(H(q)-\zeta_2-\beta_1)}-\log(\epsilon_1^{-1})\label{eq:h:inf:2}
    \end{align}

The proof appears in Appendix~\ref{app:smooth:min:entropy}. \\
\end{lemma}

From Lemma~\ref{lem:smooth:min:entropy} and Lemma~\ref{lem:smooth:min:entropy2}, we have:
\begin{align}
    H_{\infty}(\bX) &\geq \min\{H_{\infty}(\bX|\bY, G, G(\bX)),H_{\infty}(\bX|\bZ, G, G(\bX))\}\notag\\
    &\geq n\left\{\min\{H(p),H(q)\}-\max\{\zeta_1,\zeta_2\}-\beta_1\right\}\notag\\
    & \hspace*{50mm}-\log(\epsilon_1^{-1})\label{eq:ee2}
\end{align}

Now, we crucially use leftover hash lemma (Lemma~\ref{lem:glh}) to show that $nR$ nearly random bits can be extracted from $\bX$ in the form of $\text{Ext}(\bX)$ using the $2-$Universal hash function Ext (the key $\text{Ext}(\bX)$ has nearly uniform distribution). This shows that the protocol satisfies \emph{bias-based secrecy}. 

Let us fix $\epsilon_1:=2^{-n\alpha_2}$, where $\alpha_2>0$ is an arbitrary small constant.
We make the following correspondence in  Lemma~\ref{lem:glh}: $G\leftrightarrow \text{Ext}$, $W\leftrightarrow \bX$ and  $l\leftrightarrow nR$
to get the following:
\begin{align}
    \|&P_{\text{Ext}(\bX),\text{Ext}}-P_{U_l,\text{Ext}}\| \notag \\
    &\stackrel{(a)}{\leq} \frac{1}{2}\sqrt{2^{ -H_{\infty}(\bX)} 2{^{nR}}}\notag\\
    &\stackrel{(b)}{\leq} \frac{1}{2}\sqrt{2^{-n(\min\{H(p),H(q)\}-\max\{\zeta_1,\zeta_2\}-\beta_1-\alpha_2)}}\notag\\ &\hspace{45mm} .\sqrt{2^{n(\min\{H(p),H(q)\}- \beta_2)}}\notag\\
    &= \frac{1}{2}\sqrt{2^{n(\max\{\zeta_1,\zeta_2\}+\beta_1+\alpha_2-\beta_2))}}\notag\\
    &\stackrel{(c)}{\leq} 2^{-n\alpha_3}  \label{eq:sd:1}
\end{align}
where, $\alpha_3 > 0$ and $n$ is sufficiently large.
Here,\begin{enumerate}[(a)]
\item follows directly from Lemma~\ref{lem:glh}.
\item follows from (\ref{eq:ee2}) and noting that the choice of $\text{Ext}$ is random and  uniform  from ${\mathcal{E}}:\{0,1\}^n \rightarrow \{0,1\}^{nR}$ where $R:=\min\{H(p),H(q)\}- \beta_2$.
\item follows from noting that $\beta_2$ is chosen such that $\max\{\zeta_1,\zeta_2\}+\beta_1+\alpha_2-\beta_2<0$; here, we note that $\alpha_2$ is an arbitrarily chosen (small enough) constant, and $\zeta_1,\zeta_2>0$  can be made arbitrarily small for sufficiently large $n$. Thus, a choice of $\beta_2>\beta_1$ is sufficient.
\end{enumerate}
Thus, this proves that our commitment protocol satisfies bias-based secrecy (cf.~\cite[Def.~3.1]{damgard1998statistical}).  Recall from our discussion earlier (see also~\cite[Th.~4.1]{damgard1998statistical}) that bias-based secrecy under \emph{exponentially decaying} statistical distance, as in~\eqref{eq:sd:1}, implies capacity-based secrecy; hence, it follows that for $n$ sufficiently large, we have  $\max\{I(C;\view_B),I(C;V_E)\}\leq \epsilon$ and our protocol is $\epsilon-1-$concealing as well as $\epsilon-$secure.\\

[3] \underline{\emph{$\epsilon-1-$binding:}}
A commitment protocol satisfies $\epsilon-1-$bindingness if under \emph{any} behaviour of a malicicous Alice (without colluding with Eve), Bob can verify (with high probability) if Alice's revelation in the reveal phase $(\td{c},\td{\bx})$ are similar or different to it's choices in the commit phase.\\

Note that here in the $1-$privacy case, we only need to guarantee bindingness only between a malicious Alice (who \textit{does not} colludes with the wiretapper Eve) and a honest Bob. Therefore, the $1-$bindingess analysis is almost similar to the case in which the wiretapper Eve is absent and there is a one-way \textit{BSC(p)} from Alice to Bob. 

 Thus, let $\bX=\bx$ be the transmitted bit string and $\bY=\by$ be the  bit string received by Bob's over the {I-BS-BC$(p,q)$}. Alice can cheat successfully by confusing Bob in the reveal phase only if she can find two distinct bit strings $\bx'$ and $\td{\bx}$ such that (i) $\bx',\td{\bx} \in \cL(\by)$, and (ii) $\bx'$, $\td{\bx}$ pass the random hash exchange challenge (w.r.t hash functon $G(\cdot)$). The number of such strings that Alice can use to confuse Bob and that can pass the Bob's first test in the reveal phase are exponentially many in $n$; in particular, if $\cA$ denotes this set of strings that Alice can choose to reveal in the reveal phase. Then, we have
\begin{align}
|\cA|&\stackrel{}{\leq} 2^{n\eta}\label{eq:A}
\end{align}
where $\eta>0$. 

Next, we show that the probability of hash collision for any two bit strings $\bx$ and $\bx'$ in $\mathcal{A}$ is exponentially decaying in $n$, i,e., the hash challenge prevents any malicious action of Alice.
\begin{claim}\label{claim:step:2}
For $n$ sufficiently large,
\begin{align}
\mathbb{P}&\Bigg(\exists \bx\neq \bx'\in\mathcal{A}:G(\bx)=G(\bx')\Bigg)
\leq 2^{-n\beta'} \label{eq:bind:1}
\end{align}
The proof of the Claim appears in the Appendix~\ref{app:step:2}.
\end{claim}

The fact that the secuirty parameter $\beta'>0$, shows that our commitment protocol is $\epsilon-$binding.

\subsection{Achievability for I-BS-BC$(p,q)$ wiretap channel under $2-$privacy. }\label{sec:proofs:achieve:1}

 Similar to the previous case, this achievability protocol is also inspired by works in~\cite{damgaard1999possibility,crepeau2020commitment,yadav2021commitment} and just requires \textit{one round of random hash challenge} essentially to \emph{bind} Alice to her choice in the commit phase.\\

Here are the details of our protocol.  The rate $R:=\min\{H(p)+H(q)-H(p \circledast q)\}- \beta_2$, where the choice of $\beta_2>0$ is specified later. 
Let $\mathcal{G}:=\{g:\{0,1\}^n \rightarrow \{0,1\}^{n\beta_1}\}$ be a $2-$universal hash family, where $\beta_2>\beta_1>0$ is a small enough constant.
Further, let $\mathcal{E}:=\{\text{ext}:\{0,1\}^n \rightarrow \{0,1\}^{nR}\}$ be a $2-$universal hash family. \footnote{Note that  $R$ can be made arbitrarily close to $\mathbb{C}.$}

We now describe the commit and reveal phases:\\

\indent $\bullet$ \emph{Commit Phase:}  
To commit string $c\in[2^{nR}]$, the protocol proceeds as follows:\\

\noindent (C1). Given $c$, Alice sends $\bX\sim \text{Bernoulli}(1/2)$ independent and identically distributed (i.i.d.)  over the I-BS-BC$(p,q)$ wiretap channel; Bob receives $\bY$ and Eve receives $\bZ$. \\

\noindent (C2). Bob chooses a hash function $G\sim \text{Unif}\left(\mathcal{G}\right)$, and sends the description of $G$ to Alice over the noiseless channel.\\ 

\noindent (C3). Alice computes $G(\bX)$ and sends it to Bob over the noiseless channel.\\

\noindent (C4). Alice chooses an extractor function $\text{Ext}\sim\text{Unif}\left(\mathcal{E}\right)$ and sends $Q = c \oplus \text{Ext}(\bX)$ and the description of $\text{Ext}$ to Bob i.e., $(Q,\text{Ext})$ over the noiseless link.\footnote{In the following expression, operator $\oplus$ denotes component-wise XOR.}\\

\indent $\bullet$ \emph{Reveal Phase:} Alice proceeds as follows: \\

\noindent {(R1). Having received $\bY=\by$, Bob creates list $\cL(\by)$ of vectors given by:\footnote{Here the parameter $\alpha_1>0$ is chosen appropriately small.}
\begin{IEEEeqnarray*}{rCl}
\cL(\by):=\{\bx\in \{0,1\}^n: n(p -\alpha_1) \leq d_H(\bx,\by) \leq n(p +\alpha_1) \}.
\end{IEEEeqnarray*}
}

\noindent (R2). Alice announces  $(\td{c},\td{\bx})$ to Bob over the noiseless link.\\

\noindent (R3). Bob accepts $\td{c}$ if all the following three conditions are satisfied: \\
$(i)$ $\td{\bx}\in\cL(\by)$, \\
$(ii)$ $g(\tilde{\bx})=g({\bx})$, and \\
$(iii)$ $\tilde{c}=q~\oplus \text{ext}(\tilde{\bx})$. 

Else, he rejects $\td{c}$ and outputs `0'.\\

Note that the \textit{eavesdropper} Eve also has the access to the noiseless channel so it observes everything shared between Alice and Bob over the noiseless channel during the commit as well as the reveal phase.

We now analyse and prove the security guarantees under $2-$privacy in detail for the $(n,R)$-commitment scheme, defined above:

[1] \underline{\emph{$\epsilon-$sound:}} This is similar to proving the $\epsilon-$soundness for the case of $1-$privacy. For our protocol to be $\epsilon$-sound, it is sufficient to show that  $\bbP\left(\bX\not\in \cL(\bY)\right)\leq \epsilon$ when both the parties, Alice and Bob, are honest; the proof for the same follows from classic Chernoff bounds. \\

[2] \underline{\emph{$\epsilon-2-$concealing:}}
It is known that a positive rate commitment protocol is $\epsilon-2-$concealing if it satisfies the notion of \emph{capacity-based secrecy} (cf.~\cite[Def.~3.2]{damgard1998statistical}) i.e., $I(C;V_B,V_E) \leq \epsilon$, and vice versa. We use a well established relation between \emph{capacity-based secrecy} and the \emph{bias-based secrecy} (cf.~\cite[Th.~4.1]{damgard1998statistical}) to prove that our protocol is $\epsilon-2-$concealing. 

We begin by establishing the following lower bound in Lemma~\ref{lem:smooth:min:entropy22} and Lemma~\ref{lem:smooth:min:entropy2} which quantify the left-over uncertainity in $\bX$, after the information about $\bX$ is lost to colluding Bob and Eve due to the collective access of $(\bY,\bZ, G,G(\bX))$:
\begin{lemma}\label{lem:smooth:min:entropy22}
For any $\epsilon_1>0, \zeta_1>0$ and $n$ sufficiently large, 
\begin{align}
	H_{\infty}^{\epsilon_1}&(\bX|\bY, \bZ, G, G(\bX))\notag\\
	&\stackrel{}{\geq} { n(H(p)+H(q)-H(p \circledast q)-\zeta_1-\beta_1)}-\log(\epsilon_1^{-1})\label{eq:h:inf:1}
    \end{align}
    The proof appears in Appendix~\ref{app:smooth:min:entropy22}.
\end{lemma}

Now, we crucially use leftover hash lemma (Lemma~\ref{lem:glh}) to show that $nR$ nearly random bits can be extracted from $\bX$ in the form of $\text{Ext}(\bX)$ using the $2-$Universal hash function Ext (the key $\text{Ext}(\bX)$ has nearly uniform distribution). This shows that the protocol satisfies \emph{bias-based secrecy}. 

Let us fix $\epsilon_1:=2^{-n\alpha_2}$, where $\alpha_2>0$ is an arbitrary small constant.
We make the following correspondence in  Lemma~\ref{lem:glh}: $G\leftrightarrow \text{Ext}$, $W\leftrightarrow \bX$ and  $l\leftrightarrow nR$
to get the following:
\begin{align}
    \|(P_{\text{Ext}(\bX),\text{Ext}}&-P_{U_{l},\text{Ext}})\| \notag \\
     &\stackrel{(a)}{\leq} \frac{1}{2}\sqrt{2^{ -H_{\infty}(\bX)} 2{^{l}}}\notag\\
	&\stackrel{(b)}{\leq} \frac{1}{2}\sqrt{2^{ -H_{\infty}(\bX|\bY,\bZ, G(\bX),G)} 2{^{l}}}\notag\\
    &\stackrel{(c)}{\leq} \frac{1}{2}\sqrt{2^{-n(H(p)+H(q)-H(p \circledast q)-\zeta-\beta_1-\alpha_2)}}\notag\\ &\hspace{18mm} .\sqrt{2^{n(H(p)+H(q)-H(p \circledast q)- \beta_2)}}\notag\\
    &= \frac{1}{2}\sqrt{2^{n(\zeta+\beta_1-\beta_2+\alpha_2))}}\notag\\
    &\stackrel{(d)}{\leq} 2^{-n\alpha_3}  \label{eq:sd:189}
\end{align}
where, $\alpha_3 > 0$ and $n$ is sufficiently large.
Here,\begin{enumerate}[(a)]
\item follows from~Lemma~\ref{lem:glh}.
\item follows from noting that min-entropy upper bounds conditional min-entropy.
\item follows from~Lemma~\ref{lem:smooth:min:entropy} and noting that  $R:=H(p)+H(q)-H(p \circledast q)- \beta_2$ for arbitrarily small $\beta_2$.
\item follows from noting that $\beta_2$ is chosen such that $\zeta+\beta_1+\alpha_2-\beta_2<0$; here, we note that $\zeta$ and $\alpha_2$ can be made arbitrarily small for sufficiently large $n$, therefore, a choice of $\beta_2>\beta_1$ is sufficient.
\end{enumerate}

From~\eqref{eq:sd:189} and Lemma~\ref{lem:glh}, it follows that the specified commitment protocol satisfies biased-based secrecy which further implies capacity-based secrecy. Thus, our protocol satisfies $\epsilon$-concealment for sufficiently large $n$.\\

[3] \underline{\emph{$\epsilon-2-$binding:}}
A commitment protocol satisfies $\epsilon$-bindingness if under \emph{any} behaviour of colluding Alice and Eve, Bob is able to verify (with high probability) if Alice's revelation in the reveal phase $(\td{c},\td{\bx})$ are similar to it's choices in the commit phase or are different.\\

Note that for the independent broadcast wiretap channel $W_{Y,Z|X}$, we have the following decomposition $W_{Y,Z|X}=W_{Y|X}W_{Z|X}$, and thus the following markov chain $Y-X-Z$ holds. 

It implies that eventually Alice colluding with Eve doesn't helps Alice in extracting any extra information about the vector $\bY$ received by Bob i.e., $H(\bY|\bX,\bZ)=H(\bY|\bX)$. Thus, the sender Alice in the $2-$privacy case is only as powerful as in the $1-$privacy case without colluding with Eve. Thus, our bindingness analysis for the $2-$privacy case is similar to the analysis for the $1-$privacy case, in the previous subsection. This completes our analysis for $\epsilon-2-$bindingness.\\

[4] \underline{\emph{$\epsilon-$secure:}} A commitment protocol is $\epsilon-$secure if for any behaviour of the wiretapper Eve, we have $I(C;V_E) \leq \epsilon$. 

This directly holds due to the fact that our commitment protocol satisfies $\epsilon-2-$concealment. As a result, we have
\begin{align}
    I(C;V_E) &\leq I(C;V_E,V_B) \notag\\
    &\leq \epsilon
\end{align}
where $\epsilon$ is exponentially decaying in the blocklength $n$. Therefore, our commitment protocol is $\epsilon-$secure against Eve.

\section{Concluding  Remarks and Discussion}\label{sec:conclusion}

We initiated the study of \textit{wiretapped commitment} in the presence of an eavesdropper in this work. We studied the maximum commitment throughput \textit{a.k.a commitment capacity} of certain subclasses of wiretap channels by providing security guarantees under two regimes i.e., $1-$privacy - where the eavesdropper \textit{cannot} collude with any of the legitimate parties of the commitment protocol, and $2-$privacy - in which the eavesdropper can collude with (atmost) one of the legitimate (malicious) parties to affect the protocol.

Our converse bounds under the $1-$privacy regime as well as under the $2-$privacy regime hold for any wiretap channel.  
Then, under $1-$privacy we provided a matching achievability for BS-BC wiretap channels and completely characterized their $1-$privacy commitment capacity. In the $2-$privacy regime, we provided a matching achievability for I-BS-BC wiretap channel (which form a sub-class of the BS-BC wiretap channels) and thus, also completely characterized the $2-$privacy commitment capacity of the I-BS-BC wiretap channel. 
Also, it is important to note that our $1-$privacy converse proof for $R\leq \max_{P_X}H(X|Z)$ assumes the independence of the public communication and the channel output, given the channel input. While, it seems a fairly natural assumption in commitment protocols, we believe there might exist a more general converse proof which bypasses this asumption of the markov chain.

The degraded binary symmetric broadcast (D-BS-BC) wiretap channel presents an interesting challenge in the $2-$privacy regime. While the upper bound in Eq.~\eqref{eq:2:private:cap} holds, it can be shown that the bound is quite weak (we conjecture that the capacity in this case will be strictly lower than in Eq.~\eqref{eq:2:private:cap} for meaningful values of $p,q$). The key challenge for this setup resides in analysing the effect of collusion between Alice and Eve (and the accompanying commitment rate penalty). 

\section{Acknowledgements}\label{sec:ack}
A.~K.~Yadav was partially supported by the Swiss National Science Foundation (SNSF) grant N\textsuperscript{o} 211337.
M.~Mamindlapally was partially supported by the Dutch Research Council (NWO) as part of the project Divide and Quantum `D\&Q' NWA.1389.20.241 of the program `NWA-ORC'.
A.~J.~Budkuley was partially supported by Dept. of Science and Technology, Govt. of India, through grant MTR/2023/001412.

\clearpage
\balance
\bibliographystyle{IEEEtran}
\bibliography{IEEEabrv,References}
\newpage
\appendix

\subsection{Proof of Lemma~\ref{lem:H:C:XV} and Lemma~\ref{lem:H:C:XV11}}\label{app:A}
Recall that $V_B$ denotes the  view of Bob at the end of commit phase. 

Let's define\footnote{Although Bob's test $T$ is a randomized test, it can be shown that one can construct from $T$ a deterministic test with essentially the same soundness and bindingness performance. Hence, for the rest of the converse, we consider that Bob's test is a deterministic function; as such, $\td{c}$ is well defined for such a deterministic test.} 
%
$\td{c}:=\arg \max_{c\in[2^{nR}]} T(\td{c},\bX,V_B).$
%
Now, we will bound from above $\mathbb{P}(\hat{C}\neq C)$, where $\hat{C}=\hat{C}(V_B,\bX)=\td{c}$. 
As the commitment scheme is $\epsilon_n-2-$binding for the $2-$privacy case (similarly, $1-$binding for the $1-$privacy case), we know that,
\begin{align}
\mathbb{P}\left(T(\bar{c},\bar{\bX},V_B)=1 \quad\& \quad T(\hat{c},\hat{\bX},V_B)=1 \right)\leq \epsilon_n
\end{align}
for any two distinct $(\bar{c},\bar{\bX})$ and $(\hat{c},\hat{\bX})$ under \textit{any} behaviour of Alice under possible collusion with Eve (\textit{any} behaviour of Alice under no collusion with Eve, in the $1-$privacy case). 
Thus, for the given decoder, we have
\begin{align}
\mathbb{P}(\hat{C}\neq C)&=\mathbb{P}(\hat{C}=0)+\mathbb{P}(\hat{C}\neq C|C\neq 0) \notag \\
												  &\leq \epsilon_n+ \epsilon_n \notag \\
													&=2\epsilon_n. \label{eq:msd}
\end{align}
where in the penultimate inequality, the first part follows from noting that  $\mathscr{P}_n$ is $\epsilon_n-2-$binding, and the second part follows from the fact that conditioned on $\mathscr{P}_n$ being $\epsilon_n-2-$binding, the probability that $\hat{C}$ is different from $C$ is at most $\epsilon_n$ due to $\mathscr{P}_n$ being $\epsilon_n$-sound.

We now use Fano's inequality (cf.~\cite{elgamal-kim}) to upper bound the following conditional entropy.
\begin{align}
H(C|\bX,V_B,V_E)&\stackrel{(a)}{\leq} H(C|\bX,V_B)\notag\\&\stackrel{(b)}{\leq} 1+ \mathbb{P}(\hat{C}\neq C) nR \notag\\
           &\stackrel{(c)}{\leq} n\left(\frac{1}{n}+ 2\epsilon_n R \right) \notag \\
					 &\leq n\epsilon'_n 
\end{align}
where $\epsilon'_n\rightarrow 0$ as $n\rightarrow \infty$, and
\begin{enumerate}[(a)]
\item follows from noting that conditioning reduces entropy.
\item follows from the Fano's inequality (cf.~\cite{elgamal-kim}).
\item follows from Eq.~\eqref{eq:msd}.
\end{enumerate}

This completes the proof of the Lemma~\ref{lem:H:C:XV} as well as Lemma~\ref{lem:H:C:XV11}.

\subsection{Proof of Lemma~\ref{lem:smooth:min:entropy} and  Lemmma~\ref{lem:smooth:min:entropy2}\label{app:smooth:min:entropy}}
    Before we start with the proof, we recap a few well known results (without proof) which will be needed in our proof.
\begin{claim}[Min-entropy~\cite{chain1, chain2}]\label{claim:min:entropy}
 For any $0 \leq \mu,\mu',\mu_1,\mu_2 <1$ and any set of jointly distributed random variables $(X, Y, W)$, we have 

\begin{IEEEeqnarray}{rCl}
&&H_{\infty}^{\mu+\mu^{'}}(X,Y|W)-H_{\infty}^{\mu^{'}}(Y|W)\notag\\ 
&&\geq H_{\infty}^{\mu}(X|Y,W)\label{eq:min:1}\\ 
&&\geq H_{\infty}^{\mu_1}(X,Y|W)-H_0^{\mu_2}(Y|W)-\log\left[\frac{1}{\mu-\mu_1-\mu_2}\right] \label{eq:min:2}
\end{IEEEeqnarray}

\end{claim}
\begin{claim}[Max-entropy~\cite{chain1, chain2}]\label{claim:max:entropy}
For any $0 \leq \mu,\mu',\mu_1,\mu_2 <1$ and any set of jointly distributed random variables $(X, Y, W)$, we have 
\begin{IEEEeqnarray}{rCl}
&&H_{0}^{\mu+\mu^{'}}(X,Y|W)-H_{0}^{\mu^{'}}(Y|W) \notag\\
&&\leq H_{0}^{\mu}(X|Y,W)\label{eq:max:1}\\
&&\leq H_{0}^{\mu_1}(X,Y|W)-H_{\infty}^{\mu_2}(Y|W)+\log\left[\frac{1}{\mu-\mu_1-\mu_2}\right]\label{eq:max:2}
\end{IEEEeqnarray}
\end{claim}
Using the above two claims establishing a lower bound on the following smooth-min-entropy:
\begin{align}
	H_{\infty}^{\epsilon_1}&(\bX|\bY,G(\bX),G)\notag\\
	&\stackrel{(a)}{\geq} H_{\infty}(\bX,G(\bX),|\bY,G)\notag\\
	&\hspace{10mm}-H_{0}(G(\bX)|\bY,G)-\log(\epsilon_1^{-1})\notag\\
	&\stackrel{(b)}{\geq} H_{\infty}(\bX|\bY,G)+H_{\infty}(G(\bX)|\bY,G,\bX)\notag\\
	&\hspace{10mm}-H_{0}(G(\bX)|\bY,G)-\log(\epsilon_1^{-1})\notag\\
	&\stackrel{(c)}{\geq} H_{\infty}(\bX|\bY,G)\notag\\
	&\hspace{10mm}-H_{0}(G(\bX)|\bY,G)- \log(\epsilon_1^{-1})\notag\\
	&\stackrel{(d)}{=} H_{\infty}(\bX|\bY)-H_{0}(G(\bX)|\bY,G)- \log(\epsilon_1^{-1})\notag\\
	&\stackrel{(e)}{\geq}  (H(\bX|\bY)-\zeta_1)-H_{0}(G(\bX)|\bY,G)-\log(\epsilon_1^{-1})\notag\\
	&\stackrel{(f)}{\geq} { n(H(p)-\zeta_1)}-n\beta_1-\log(\epsilon_1^{-1})\notag\\
	&\stackrel{}{=} { n(H(p)-\zeta_1-\beta_1)}-\log(\epsilon_1^{-1})\label{eq:h:inf:1}
    \end{align}
where we have
\begin{enumerate}[(a)]
\item from the chain rule for smooth min-entropy; see Claim~\ref{claim:min:entropy} and substitute $\mu=\epsilon_1$, $\mu_1=0$ and $\mu_2=0$ in Eq.~\eqref{eq:min:2}.
\item from the chain rule for min-entropy; see Claim~\ref{claim:min:entropy} and and substitute $\mu=0$ and $\mu'=0$ in Eq.~\eqref{eq:min:1}.
\item from the fact that $G(\bX)$ is a deterministic function of $G$ and $\bX$. 
\item by the Markov chain  $\bX - \bY - G$.
\item from~\cite[Th.~1]{nascimento-barros-t-it2008} which allows us to lower bound $H_{\infty}(\bX|\bY)$ in terms of $H(\bX|\bY)$
\item by noting that the effective channel from Alice to Bob is a \textit{BSC(p)}, and from definition of max-entropy (also noting that the  range of $G$ is $\{0,1\}^{n\beta_1}$).
\end{enumerate}
\begin{remark}
    The proof for Lemma~\ref{lem:smooth:min:entropy2} follows similarly. Note that in this case, the following markov chain $\bX-\bZ-G$ exists. Additionally, we have $H_{\infty}(\bX|\bZ) \geq H(\bX|\bZ)-\zeta_2 = H(q)-\zeta_2$, for arbitrarily small constant $\zeta_2>0$.
\end{remark}

\subsection{Proof of Claim~\ref{claim:step:2}}\label{app:step:2}
Recall that $G\sim\text{Unif}\left(\mathcal{G}\right)$, where $\mathcal{G}=\{g:\{0,1\}^n \rightarrow \{0,1\}^{n\beta_1}\}$. Therefore for any $\bx,\bx'$ $\in$ $\{0,1\}^n$, we have 
\begin{align}
    \mathbb{P}\Bigg(G(\bx)=G(\bx')\Bigg) \leq \frac{1}{2^{n\beta_1}} \label{eq:hash}
\end{align}
Now,
\begin{align}
\mathbb{P}&\Bigg(\exists \bx\neq \bx'\in\mathcal{A}:G(\bx)=G(\bx')\Bigg)\notag\\
&\stackrel{(a)}{\leq} {|\mathcal{A}|\choose {2}} \mathbb{P}\left(G(\bx)=G(\bx')\right) \notag\\
&\stackrel{(b)}{\leq} {{2^{n\eta}}\choose {2}}2^{-n\beta_2} \notag\\
&\stackrel{}{<} 2^{2n\eta}2^{-n\beta_1} \notag\\
&\stackrel{(c)}{\leq}2^{-n\beta'}
\end{align}
where $(a)$ follows from the definition of $\mathcal{A}$, and using the union bound (on distinct pairs of vectors in $\mathcal{A}$); we get $(b)$ from the definition of $\mathcal{G}$. Further, $(c)$ follows from the fact that $\beta_1$ is chosen such that $\beta':=\beta_1-2\eta >0$.
This completes the proof of the claim.

\subsection{Proof of Lemma~\ref{lem:smooth:min:entropy22} \label{app:smooth:min:entropy22}}
Using the Claim~\ref{claim:min:entropy} and Claim~\ref{claim:max:entropy}, we  establish a lower bound on the following smooth-min-entropy:
\begin{align}
	H_{\infty}^{\epsilon_1}&(\bX|\bY,\bZ,G, G(\bX))\notag\\
	&\stackrel{(a)}{\geq} H_{\infty}(\bX,G(\bX),|\bY,\bZ,G)\notag\\
	&\hspace{10mm}-H_{0}(G(\bX)|\bY,\bZ,G)-\log(\epsilon_1^{-1})\notag\\
	&\stackrel{(b)}{\geq} H_{\infty}(\bX|\bY,\bZ,G)+H_{\infty}(G(\bX)|\bY,\bZ,G,\bX)\notag\\
	&\hspace{10mm}-H_{0}(G(\bX)|\bY,\bZ,G)-\log(\epsilon_1^{-1})\notag\\
	&\stackrel{(c)}{\geq} H_{\infty}(\bX|\bY,\bZ,G)\notag\\
	&\hspace{10mm}-H_{0}(G(\bX)|\bY,G)- \log(\epsilon_1^{-1})\notag\\
	&\stackrel{(d)}{=} H_{\infty}(\bX|\bY,\bZ)-H_{0}(G(\bX)|\bY,\bZ,G)- \log(\epsilon_1^{-1})\notag\\
	&\stackrel{(e)}{\geq}  (H(\bX|\bY,\bZ)-\zeta_1)-H_{0}(G(\bX)|\bY,\bZ,G)-\log(\epsilon_1^{-1})\notag\\
	&\stackrel{(f)}{\geq} { n(H(p)+H(q)-H(p \circledast q)-\zeta_1)}-n\beta_1-\log(\epsilon_1^{-1})\notag\\
	&\stackrel{}{=} { n(H(p)+H(q)-H(p \circledast q)-\zeta_1-\beta_1)}-\log(\epsilon_1^{-1})\label{eq:h:inf:1}
    \end{align}
where we have
\begin{enumerate}[(a)]
\item from the chain rule for smooth min-entropy; see Claim~\ref{claim:min:entropy} and substitute $\mu=\epsilon_1$, $\mu_1=0$ and $\mu_2=0$ in Eq.~\eqref{eq:min:2}.
\item from the chain rule for min-entropy; see Claim~\ref{claim:min:entropy} and and substitute $\mu=0$ and $\mu'=0$ in Eq.~\eqref{eq:min:1}.
\item from the fact that $G(\bX)$ is a deterministic function of $G$ and $\bX$. 
\item by the Markov chain  $\bX - (\bY,\bZ)- G$.
\item from~\cite[Th.~1]{nascimento-barros-t-it2008} which allows us to lower bound $H_{\infty}(\bX|\bY,\bZ)$ in terms of $H(\bX|\bY,\bZ)$.
\item by noting that $H(\bX|\bY,\bZ)=H(p)+H(q)-H(p \circledast q)$, and from definition of max-entropy (also noting that the  range of $G$ is $\{0,1\}^{n\beta_1}$).
\end{enumerate}


\end{document}